\documentstyle[twocolumn,aps,psfig]{revtex}

\newlength{\FigWidth}
\begin{document}
\draft
\preprint{HEP/123-qed}
\title{Spontaneous deformation of the Fermi surface
due to strong correlation
in the two-dimensional $t$-$J$ model}

\author{
A. Himeda$^*$ and M. Ogata
\thanks{From April 2000, Department of Physics, 
University of Tokyo, Hongo, Bunkyo-ku, 
Tokyo 113-0033, Japan}
}

\address{
Department of Basic Science, Graduate School of Arts and Sciences, \\
University of Tokyo, Komaba, Meguro-ku, Tokyo 153-8902, Japan
}

\date{\today}

\maketitle
\begin{abstract}
Fermi surface of the two-dimensional $t$-$J$ model 
is studied using the variational Monte Carlo method.
We study the Gutzwiller projected $d$-wave superconducting state
with an additional variational parameter $t'_{\rm v}$
corresponding to the next-nearest neighbor hopping term.
It is  found that the finite $t'_{\rm v}<0$
gives the lowest variational energy
in the wide range of hole-doping rates.
The obtained momentum distribution function shows that
the Fermi surface deforms spontaneously.
It is also shown that the van Hove singularity 
is always located very close to the Fermi energy.
Using the Gutzwiller approximation,
we show that this spontaneous deformation 
is due to the Gutzwiller projection operator
or the strong correlation.
\end{abstract}
\narrowtext
\newcommand{\mybf}[1]{\mbox{\boldmath $#1$}}
\newcommand{\lsim}
 {\ \raise.35ex\hbox{$<$}\kern-0.75em\lower.5ex\hbox{$\sim$}\ }
\newcommand{\gsim}
 {\ \raise.35ex\hbox{$>$}\kern-0.75em\lower.5ex\hbox{$\sim$}\ }

The effect of strong correlation is 
one of the most important issues for understanding
the high-$T_c$ superconductivity (SC).
Among various anomalous electronic properties,
the experiments of angle resolved photoemission spectroscopy (ARPES)
have revealed that a flat band around ($\pi$, 0) and (0, $\pi$)
is pinned just below the Fermi energy.\cite{dess,marshall,ino}
This phenomenon is unexpected in the band calculations
and it is considered to be closely related to the 
opening of the pseudogap on the Fermi surface (FS)
\cite{yasuoka,rossat,norman},
which is also an extraordinary feature in high-$T_c$ cuprates.
This anomalous nature of the FS will be the direct evidence
for the non-Fermi liquid behavior.
It is thus an interesting issue to study the FS
in the presence of strong correlation.

The effect of the flat band and the geometry of the FS
can be taken into account by using
the $t$-$t'$-$J$ model or the $t$-$t'$-$U$ Hubbard model
in which the next-nearest neighbor hopping term
$t'$ is introduced as a fitting parameter.\cite{tanamoto}
If one chooses $t'<0$, the FS centered at ($\pi$, $\pi$)
observed experimentally\cite{ino,aebi}
can be reproduced in the tight-binding model.
However high temperature expansion studies
on the momentum distribution function 
for the $t$-$J$ model\cite{putikka}
have shown that the FS
is similar to that with $t'<0$ even though 
the $t'$-term is absent in the Hamiltonian.
On the other hand, the conventional mean-field theories,
such as slave-boson theory, simply give the FS with $t'=0$.
Therefore the strong correlation which is not included
in the mean-field theories will be 
the origin of the change of the FS geometry.

Here we study this problem from a different point of view.
Since the calculation in Ref.\cite{putikka} is carried out
in the high temperature region, it is not clear
whether or not the FS deforms down to zero temperature.
To study the FS of the ground state is generally very difficult.
The exact diagonalization study of small clusters does not 
give enough resolution in the $\mybf{k}$ space. 
The quantum Monte Carlo simulations
have been often useless for the two-dimensional $t$-$J$ model
because of the minus sign problem.
Therefore we use the variational Monte Carlo (VMC) method
in this paper, which is free from the limitation of 
the system size as well as from the sign problem.
The VMC method treats exactly the constraints 
of no doubly occupied sites and gives accurate estimates
of the expectation values such as the variational energies
and the momentum distribution functions.

Although it is a variational theory,
the VMC method is powerful to see whether some kind of 
symmetry breaking takes place or not.
In this paper we examine the Gutzwiller-projected
$d$-wave superconducting state 
which contains an additional variational parameter $t'_{\rm v}$
corresponding to the next-nearest neighbor hopping term.
We can safely discuss the relative energy difference
between the variational states with and without $t'_{\rm v}$,
although the absolute values of the variational energies
can still be lowered.
We find that the wave function with $t'_{\rm v}\sim -0.1$
has the lowest variational energy
even though the Hamiltonian does not contain $t'$-term.
This means that the FS deforms {\it spontaneously}.
The momentum distribution function $n(\mybf{k})$
calculated in the optimized wave function 
is consistent with that in the high temperature expansion.
Our method gives an independent and complementary support
of the result that
the deformation of the FS is a distinctive feature
of strongly correlated electron systems.

In addition to this, we can identify the physical origin
of the FS deformation in our variational approach. 
We show the relation 
between the energy gain and the van Hove singularity.
It has been argued that a remarkable enhancement 
of SC correlation is achieved if the van Hove singularity
is close to the Fermi energy.\cite{feiner,yamaji}
Our results show some similarity to this picture.
Furthermore, by comparing the obtained results
with the Gutzwiller approximation,
we can see that the finite $t'_{\rm v}$
is caused solely by the Gutzwiller projection.

We use the two-dimensional $t$-$J$ model on a square lattice,
\begin{equation}
\label{hamil}
 H=-t\sum_{\langle ij\rangle \sigma}P_{\rm G}
   (c_{i\sigma}^{\dagger}c_{j\sigma}+h.c.)P_{\rm G}
   +J\sum_{\langle ij\rangle}\mybf{S}_i\cdot\mybf{S}_j   ,
\end{equation}
where $\langle ij\rangle$ represents the sum over the nearest-neighbor sites.
$c_{i\sigma}^{\dagger}$ ($c_{i\sigma}$)
is a creation (annihilation) operator of 
$\sigma$ ($\uparrow$ or $\downarrow$) electron at $i$-site 
and $\mybf{S}_i=c_{i\alpha}^{\dagger}(\frac{1}{2}
\mybf{\sigma})_{\alpha\beta}c_{i\beta}$.
The Gutzwiller's projection operator $P_{\rm G}$ is defined as
$P_{\rm G}=\Pi_i(1-\hat{n}_{i\uparrow}\hat{n}_{i\downarrow})$,
which prohibits the doubly occupied sites.
We set $J/t=0.3$.

We use a Gutzwiller-projected mean-field type wave function
$P_{\rm G}P_{N_{\rm e}}|\phi_0\rangle$ as a trial state
with fixing the number of electrons $N_{\rm e}$ through $P_{N_{\rm e}}$.
The state is written as
\begin{eqnarray}
 P_{\rm G}P_{N_{\rm e}}|\phi_0\rangle 
  &=&P_{\rm G}P_{N_{\rm e}}\prod_k(u_k+v_kc_{k\uparrow}^{\dagger}
    c_{-k\downarrow}^{\dagger})|0\rangle   \nonumber \\
  &=&P_{\rm G}P_{N_{\rm e}}\prod_k u_k \exp\left[\sum_k \frac{v_k}{u_k}
      c_{k\uparrow}^{\dagger}c_{-k\downarrow}^{\dagger}\right]|0\rangle
                      \nonumber \\
  &=& P_{\rm G}P_{N_{\rm e}}\prod_k u_k \exp\left[\sum_{ij}a_{ij}
      c_{i\uparrow}^{\dagger}c_{j\downarrow}^{\dagger}\right]|0\rangle 
                      \nonumber  \\
  &=& P_{\rm G}\prod_k u_k \frac{1}{(N_{\rm e}/2)!}
     \left(\sum_{ij}a_{ij}c_{i\uparrow}^{\dagger}
       c_{i\downarrow}^{\dagger}\right)^{N_{\rm e}/2}
        |0\rangle ,
\end{eqnarray}
where $v_k/u_k=\Delta_k/
(\epsilon _k-\mu+\sqrt{(\epsilon _k-\mu)^2+\Delta_k^2})$
and $a_{ij}$ is a Fourier transform of $v_k/u_k$.

Usually $\epsilon_k$ is chosen to be $\epsilon_k=-2(\cos k_x+\cos k_y)$,
which is in accordance with the Hamiltonian (\ref{hamil}).
However in this paper, 
we introduce an additional variational parameter $t'_{\rm v}$
which changes the FS of the variational state.
We assume
$\epsilon _k$ and $\Delta_k$ as
\begin{eqnarray}
 \epsilon _k&=&-2(\cos k_x+\cos k_y)-4t'_{\rm v}\cos k_x \cos k_y , 
                       \nonumber \\
 \Delta_k&=& 2\Delta_d (\cos k_x-\cos k_y).
\end{eqnarray}
The present wave function contains
three variational parameters $t'_{\rm v}$, $\mu$ and $\Delta_d$.

At first, using the above wave function we calculate the variational energy 
$E_{\rm var}$ of the Hamiltonian (\ref{hamil})
\begin{equation}
 E_{\rm var}=\frac{\langle \phi_0|P_{N_{\rm e}}P_{\rm G} 
              H P_{\rm G}P_{N_{\rm e}}| \phi_0 \rangle}
              {\langle \phi_0|P_{N_{\rm e}}P_{\rm G} 
               P_{\rm G}P_{N_{\rm e}}| \phi_0 \rangle},
\end{equation}
by means of the VMC method.
The distribution of the wave vectors $\mybf{k}$ is
determined in the periodic boundary conditions in the $x$ direction
and in the antiperiodic ones in the $y$ direction
so as to avoid the gap node of $d$-wave superconductivity.
Although the results in the 10$\times$10 square lattice 
are mainly shown in the following,
we also calculate larger sizes up to $20\times 20$
to investigate the size dependence.
\begin{figure}
\psfig{file=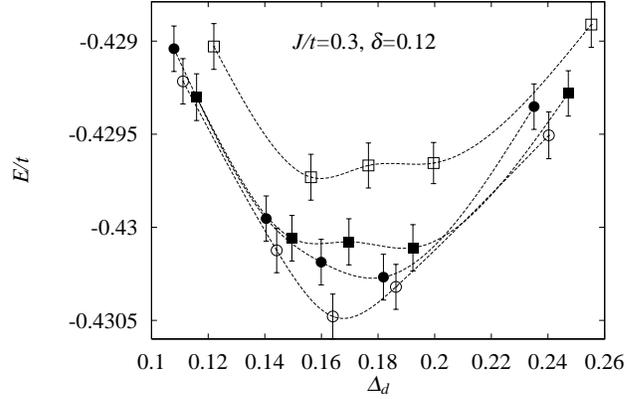,width=\FigWidth}
\caption{$\Delta_d$ dependence of $E_{\rm var}/t$ 
for $t'_{\rm v}=0$ (open squares), $-0.05$ (filled squares),
$-0.1$ (open circles) and $-0.15$ (filled circles).
Each point has been minimized with respect to $\mu$.
System size is 10$\times$10 and the number of 
Monte Carlo samples is $2.5\times 10^5$.}
\label{fig1}
\end{figure}

Figure \ref{fig1} shows the $\Delta_d$ dependence of $E_{\rm var}$
for various values of $t'_{\rm v}$ at the doping rate $\delta=0.12$.
Apparently $t'_{\rm v}\sim -0.1$ gives the lowest variational energy.
Since the Hamiltonian does not contain
next-nearest neighbor hopping terms,
the present result means 
that the shape of the FS of the ground state
is different from that of the non-interacting Hamiltonian.
We have also checked that, 
if the Hamiltonian has the next-nearest neighbor hopping term $t'$,
the optimized variational state has $t'_{\rm v}$,
whose amplitude is larger than $t'$.

\begin{figure}
\psfig{file=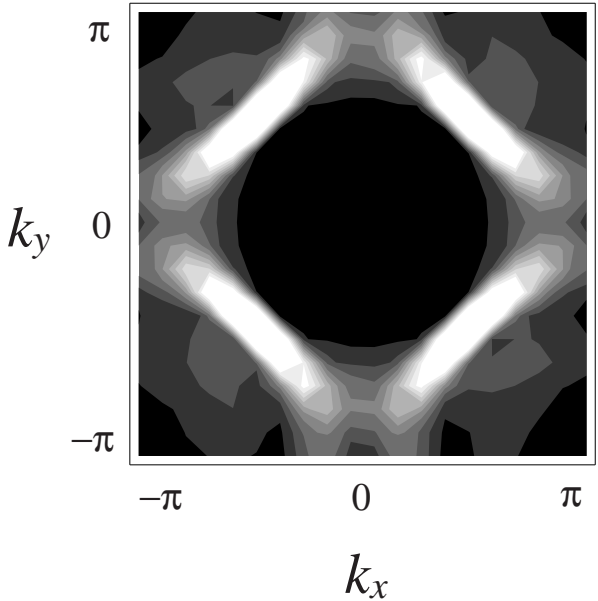,width=\FigWidth}
\caption{Full Brillouin zone plot of $|\nabla_k n(k)|$
for the doping rate $\delta =0.12$ and $J/t=0.3$.
The areas with large values of $|\nabla_k n(k)|$
are highlighted. Note that the FS
is deformed even though the Hamiltonian does not contain
the next-nearest neighbor hopping terms.
The smearing near ($\pi$, 0) and (0, $\pi$) 
is due to the presence of the $d$-wave SC order parameters.}
\label{fig2}
\end{figure}

The most significant effect of this result appears 
in the shape of momentum distribution functions.
Figure \ref{fig2} shows a contour map 
of the gradient of the momentum distribution function
$|\nabla_k n(\mybf{k})|$ for $t'_{\rm v}=-0.1$
calculated on the 20$\times$20 square lattice.
Although we have used the optimized variational parameters
$\Delta_d$ and $\mu$ on the 10$\times$10 lattice,
it is justified because their size dependeces are negligible.
Brighter areas in Fig.\ \ref{fig2} 
correspond to the momentum $\mybf{k}$ 
with larger values of  $|\nabla_k n(\mybf{k})|$.
Although we cannot specify exactly the location of the FS
due to the $d$-wave SC gap, we expect that
the FS lies close to the area
where $|\nabla_k n(\mybf{k})|$ is large.

Our result of momentum distribution function is similar to 
that obtained in high temperature expansion  
by Putikka {\it et al}.\cite{putikka}
Since we take an opposite approach to high temperature studies,
i.e., in the zero temperature,
it is confirmed that the FS shown in Fig.\ \ref{fig2}
is an intrinsic feature of the $t$-$J$ model.
Note here that the smearing of the FS around ($\pi$, 0) 
in our calculation is due to the $d$-wave SC gap.
This suggests that the similar smearing 
observed in Ref.\cite{putikka} 
may be due to the pseudogap  with $d$-wave symmetry,
in addition to the smearing due to finite temperature.

For the wide range of doping $\delta=0.04\sim 0.20$,
we find that $E_{\rm var}$ is minimized around 
$t'_{\rm v}\sim -0.1$ and the chemical potential 
$\mu \sim -0.5\pm 0.05$.
Because of the insensitiveness of $\mu$ as a function of doping,
the area of the momentum space enclosed 
by the FS is also insensitive to the doping rate.
This result supports the violation of the Luttinger theorem
suggested by Putikka {\it et al}.\cite{putikka} 
Actually $n(\mybf{k})$ at the doping $\delta=0.2$ 
in Ref.\cite{putikka} is very close to 
our results in Fig.\ \ref{fig2} at $\delta=0.12$.

\begin{figure}
\psfig{file=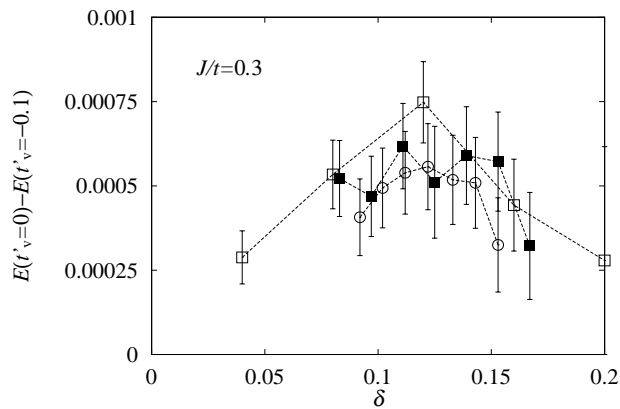,width=\FigWidth}
\caption{Doping dependence of the energy gain
at $t'_{\rm v}=-0.1$ compared with $t'_{\rm v}=0$. The system sizes are
$10\times 10$ (open squares), $12\times 12$ (filled squares) and
$14\times 14$ (open circles). The number of samples is more than 10$^5$.
}
\label{fig3}
\end{figure}

Figure \ref{fig3} shows 
the energy difference between the value at $t'_{\rm v}=0$ 
and at $t'_{\rm v}=-0.1$, i.e. the energy gain due to the finite
$t'_{\rm v}$, for several system sizes.
Although the Monte Carlo results scatter a little,
there is apparently a tendency 
that the energy gain due to the finite $t'_{\rm v}$ 
becomes maximum around $\delta=0.12$.
As we increase the system size, 
the energy gain slightly decreases,
but it will remain finite in the thermodynamic limit.

Let us discuss here the relation between the energy gain
and the van Hove singularity.
For the optimized value $t'_{\rm v}\sim -0.1$,
$\epsilon_k$ at $\mybf{k}=(\pi,0)$ becomes $-0.4$.
On the other hand, the optimized chemical potential $\mu$ 
is around $-0.5$.
This means that the suddle point 
or the position of the flat band near $\mybf{k}=(\pi ,0)$ 
is very close to the chemical potential.
Since the $d$-wave SC gap has a maximum at $(\pi,0)$,
the enhancement of the density of states near
the Fermi energy due to the van Hove singularity
causes the energy gain.
Actually, if we assume $\Delta_d=0$,
the lowest energy is achieved at $t_{\rm v}'=0$.
This indicates that the FS deforms itself
so as to fix the van Hove singularity to the chemical potential
in the presence of the $d$-wave SC gap.
This looks consistent with the mechanism 
of SC due to the van Hove singularity.\cite{feiner,yamaji}

If we use a Hamiltonian $\tilde{H}_{t{\mbox -}J}$ 
without projection operator and the mean-field wave function
$|\phi_0\rangle$, it is apparent that the variational energy
$\langle \phi_0|\tilde{H}_{t{\mbox -}J}|\phi_0\rangle/
\langle \phi_0|\phi_0\rangle$
is minimized at $t'_{\rm v}=0$.
Therefore the energy gain due to the non-zero value of $t'_{\rm v}$
is solely from the Gutzwiller's projection operator.
In order to clarify the effect of the projection,
we examine the Gutzwiller approximation\cite{ZRGS}, 
in which the effect of constraints are taken into account
by statistical weighting factors.
For the $t$-$J$ model, we have
\begin{equation}
   \frac{\langle\phi_0|P_{\rm G}c_{i\sigma}^{\dagger}c_{j\sigma}
         P_{\rm G}|\phi_0\rangle}
      {\langle\phi_0|P_{\rm G}P_{\rm G}|\phi_0\rangle}
 = g_t\langle\phi_0|c_{i\sigma}^{\dagger}c_{j\sigma}|\phi_0\rangle
 =g_t \langle c_{i\sigma}^{\dagger}c_{j\sigma}\rangle_0
\end{equation}
and 
\begin{equation}
   \frac{\langle\phi_0|P_{\rm G}\mybf{S}_i\cdot\mybf{S}_j
         P_{\rm G}|\phi_0\rangle}
      {\langle\phi_0|P_{\rm G}P_{\rm G}|\phi_0\rangle}
 = g_s\langle\phi_0|\mybf{S}_i\cdot\mybf{S}_j|\phi_0\rangle
 = g_s \langle \mybf{S}_i\cdot\mybf{S}_j\rangle_0  ,
\end{equation}
where $g_t$ and $g_s$ are the renormalization factors
due to the projection.
In the simplest Gutzwiller approximation,
$g_t$ and $g_s$ are constant,
i.e., $g_t=2\delta/(1+\delta)$ and $g_s=4/(1+\delta)^2$.
\cite{ZRGS}
In this case, the Gutzwiller projection does not alter 
the mean-field results. However it was recently shown
that the dependence of the renormalization factors
on the expectation values, such as $\chi 
=\langle c_{i\sigma}^{\dagger}c_{j\sigma}\rangle_0$
and $\Delta=\langle c_{i\uparrow}^{\dagger}c_{j\downarrow}^{\dagger}
\rangle_0$,
plays a crucial role in evaluating the variational energies.
\cite{hsu,sigrist,himeda,ogata}
If we use this Gutzwiller approximation, we can show that
\begin{equation}
  \delta E_{\rm var} \propto \frac{\partial \langle H \rangle}
        {\partial\chi '}\delta t_{\rm v}'+8Nt_{\rm v}'\delta t_{\rm v}'
           \label{ga}
\end{equation}
where $\chi '=\langle c_{i\sigma}^{\dagger}c_{j\sigma}\rangle_0$
with $(ij)$ being the next-nearest neighbor sites and
\begin{equation}
  \frac{\partial \langle H \rangle}{\partial\chi '}
  \equiv -t \frac{\partial g_t}{\partial \chi '}
        \sum_{\langle ij\rangle \sigma}
       (\langle c_{i\sigma}^{\dagger}c_{j\sigma}\rangle_0+c.c.)
      +J \frac{\partial g_s}{\partial \chi '}
        \sum_{\langle ij\rangle}
        \langle \mybf{S}_i\cdot\mybf{S}_j\rangle_0 .
\end{equation}
The first term on the r.h.s. of eq.\ (\ref{ga}) 
is linear with respect to $t_{\rm v}'$ so that
$E_{\rm var}$ is minimized at a finite value of $t_{\rm v}'$
which satisfies
\begin{equation}
 t_{\rm v}'=-\frac{1}{8N}\frac{\partial \langle H\rangle}
            {\partial \chi '} .
\end{equation}
Apparently the renormalization factors $g_t$ and $g_s$
due to the projection operator and their nonlinear dependence
on $\chi '$ are the origin 
of the spontaneous deformation of the FS.
These phenomena cannot be found in the mean-field theories.
The explicit calculations will be published elsewhere.

In summary, we investigated the shape of the FS
in the two-dimensional $t$-$J$ model by means of the VMC calculation
introducing an additional variational parameter $t'_{\rm v}$.
We found that the variational energy is minimized around
$t'_{\rm v}\sim -0.1$ for various doping rates.
The system size dependence indicate this effect is realized 
even in the thermodynamic limit.
The magnitude of the energy gain is large enough
compared with other VMC studies.
For example, 
the energy difference 
between the pure $d$-wave SC phase and the coexistent phase
of AF and $d$-wave SC is comparable to the present energy gain
at the doping rate $\delta \sim 0.08$.\cite{himeda,giam}
Then we have clarified the origin of the energy gain 
by examining the van Hove singularity and 
the effect of the projection 
using the Gutzwiller approximation.
Combining our results at zero temperature 
and those in high temperature expansion,
we consider that the FS deformation is 
the most significant phenomenon 
in the presence of strong correlation.

Since the nesting property of the FS becomes worse
in the present wave function than in the original
$t$-$J$ model, the coexistence of AF and $d$-wave SC
near half-filling\cite{himeda,giam} will be suppressed.
Alternatively we expect some incommensurate AF
correlations which were unexpected in the $t$-$J$ model.
This is presumably related to the stripe state 
observed experimentally.

\end{document}